\documentclass[twocolumn]{aastex631}
\usepackage{amssymb,amsmath}
\usepackage{graphicx}
\usepackage{natbib}
\usepackage{epstopdf}
\usepackage{hyperref}
\usepackage{bm}
\usepackage{xcolor}
\usepackage{xspace}

\newcommand\codename[1]{\textsc{#1}\xspace}
\newcommand{\parthenon}{\codename{Parthenon}\xspace}
\newcommand{\athena}{\codename{Athena++}\xspace}
\newcommand{\kathena}{\codename{K-Athena}\xspace}
\newcommand{\flash}{\codename{Flash}\xspace}
\newcommand{\kokkos}{\codename{Kokkos}\xspace}
\def\athenapk/{\codename{AthenaPK}}
\newcommand{\bhat}{\mathbf{\hat b}}

\def\k{\mathop{\rm k}\nolimits} 
 
\def\G{\mathop{\rm G}\nolimits} 
\def\g{\mathop{\rm g}\nolimits}

\def\p3m{P${}^3$M} 
\def\ap3m{AP${}^3$M} \def\-{{\em{---}}}

\def\tcc{t_{\rm cc}}

\def\gsim{\;\rlap{\lower 2.5pt
\hbox{$\sim$}}\raise 1.5pt\hbox{$>$}\;}
\def\lsim{\;\rlap{\lower 2.5pt
\hbox{$\sim$}}\raise 1.5pt\hbox{$<$}\;}

\newcommand{\be}{\begin{equation}} \newcommand{\ba}{\begin{eqnarray}}
\newcommand{\ee}{\end{equation}} \newcommand{\ea}{\end{eqnarray}}

 \newcommand{\bi}{\begin{itemize}}
\newcommand{\ei}{\end{itemize}} 
\newcommand{\K}{\,\textrm{K}}

\def\lesssim{\mathrel{\hbox{\rlap{\hbox{\lower4pt\hbox{$\sim$}}}\hbox{$<$}}}}
\def\gtrsim{\mathrel{\hbox{\rlap{\hbox{\lower4pt\hbox{$\sim$}}}\hbox{$>$}}}}

\begin{document}

\title{The Launching of Cold Clouds by Galaxy Outflows V: \\ The Role of Anisotropic Thermal Conduction}

\author{Marcus Br\"uggen}
\affiliation{Universit\"at Hamburg, Hamburger Sternwarte, Gojenbergsweg 112, 21029, Hamburg, Germany}

\author{Evan Scannapieco}, 
\affiliation{School of Earth and Space Exploration, Arizona State University, P.O. Box 871404, Tempe, AZ, 85287-1404}

\author{Philipp Grete}
\affiliation{Universit\"at Hamburg, Hamburger Sternwarte, Gojenbergsweg 112, 21029, Hamburg, Germany}

\begin{abstract}

Motivated by observations of multiphase galaxy outflows, we explore the impact of isotropic and anisotropic electron thermal conduction on the evolution of radiatively-cooled, cold clouds embedded in hot, magnetized winds. Using the adaptive mesh refinement code \athenapk/, we conduct simulations of clouds impacted by supersonic and transonic flows with magnetic fields initially aligned parallel and perpendicular to the flow direction. In cases with isotropic thermal conduction, an evaporative wind forms, stabilizing against instabilities and leading to a mass loss rate that matches the hydrodynamic case. In anisotropic cases, the impact of conduction is more limited and strongly dependent on the field orientation. In runs with initially perpendicular fields, the field lines are folded back into the tail, strongly limiting conduction, but magnetic fields act to dampen instabilities and slow the stretching of the cloud in the flow direction. In the parallel case, anisotropic conduction aids cloud survival by forming a radiative wind near the front of the cloud, which suppresses instabilities and reduces mass loss. In all cases, anisotropic conduction has a minimal impact on the acceleration of the cloud.

\end{abstract}

\section{Introduction}

All astrophysical plasmas are magnetized. Magnetic fields are essential in determining the rotation structure of the Sun and other stars \citep{2012A&A...537A.146E}, and the heating of their X-ray emitting coronae \citep{2004psci.book.....A}. Magnetic fields also play an essential role in driving accretion in the disks surrounding active black holes \citep{1977MNRAS.179..433B,1991ApJ...376..214B} and in powering the relativistic jets that are often observed to arise from them \citep{2011MNRAS.418L..79T}. In the circumgalactic medium, observations uncover magnetic field strengths of several $\mu$G \citep{2023A&A...670L..23H}, which are likely to play a significant role in the accretion history of galaxies \citep[]{Buie_2022,2023arXiv230110253F}. 

Likewise, magnetic fields are likely to play a key role in the ejection of material from starbursting galaxies. Such outflows are made up of a complex multiphase medium which ranges from $\approx10^7-10^8$ K plasma \citep[e.g.][]{1999ApJ...513..156M, 2009ApJ...697.2030S}, to $\approx10^4$ K material observed at optical and near UV wavelengths \citep[e.g.][]{2001ApJ...554..981P, 2007ApJ...663L..77T, 2012ApJ...760..127M, 2012ApJS..203....3S}, to $10-10^3$ K molecular gas \citep[e.g.][]{2002ApJ...580L..21W, 2013Natur.499..450B}. The coexistence of these different phases is not easy to explain, as the ram pressure acceleration of cold clouds in hot media is extremely inefficient. 

In the purely hydrodynamic case, studies indicate that cold clouds accelerated by a fast-moving ambient fluid disintegrate on the so-called cloud crushing time, which is the approximate time it takes for a shock in the cloud to travel a distance equal to the cloud radius, $R_{\rm c}.$  This is given by $\tcc = \chi_0^{1/2} R_{\rm c}/v_{\rm h}$, where $v_{\rm h}$ is the velocity of the hot ambient medium, and $\chi_0$ is the initial ratio of internal to external density \citep[e.g.][]{1994ApJ...420..213K,1995ApJ...454..172X,2009ApJ...703..330C,2019MNRAS.486.4526B}. 

In \cite{2015ApJ...805..158S} (hereafter Paper I), we used the FLASH code \citep{2000ApJS..131..273F} to carry out a suite of adaptive-grid hydrodynamical simulations that included radiative cooling and spanned the parameter space relevant for galaxy outflows. These simulations showed that when the hot medium exterior medium is supersonic, the Mach cone that forms around the cold-cloud suppresses shear instabilities and extends the cloud lifetime. In this case, clouds can be advected a distance of about 40 times the cloud radius before breaking up \citep[see also][]{2017ApJ...834..144S}. Other related work includes \cite{2022MNRAS.511..859G} who studied the survival of cooling clouds in a turbulent medium, which can also lead to the growth of clouds via condensation.

Cold-cloud evolution is complicated by the presence of thermal conduction, which is mediated by electrons from the hot surrounding medium that penetrate into the cold cloud, leading to evaporation.  This was studied first analytically in the case of a stationary cloud embedded in a hot plasma without \citep{1977ApJ...211..135C} and with \citep{1990ApJ...358..392M,1990ApJ...358..375B} radiative cooling. Later, simulations were able to address conduction in the case of a cloud embedded in a moving ambient medium \citep{2007A&A...472..141V, 2005MNRAS.362..626M}. In \cite{2016ApJ...822...31B} (hereafter Paper II), we included electron thermal conduction over the range of parameter space relevant for galaxy outflows and found that cloud disruption is delayed by the presence of an evaporative layer, which compresses the cloud and protects it from shredding by the exterior flow. On the other hand, evaporation leads to high cloud densities, such that the cold material is only accelerated to a small fraction of the ambient velocity, \citep[see also][]{2016MNRAS.462.4157A,2021MNRAS.502.1263K}, in conflict with observations of cold gas in outflowing starbursts \citep[e.g.][]{1999ApJ...513..156M,2001ApJ...554..981P,2002ApJ...580L..21W,2007ApJ...663L..77T, Westmoquette2011,Erb2012,2012ApJ...760..127M,Veilleux2005}.

However, these simulations assumed that conduction occurs at the full Spitzer value given by electron-ion collisions and that this value is independent of direction. In a magnetized medium, neither of these assumptions will be true. While electrons can stream freely along field lines, in the direction perpendicular to the field, they are constrained to follow circular Larmor orbits, radii that are much smaller than the mean free path for electron-ion scattering, even for astrophysical plasmas with relatively low magnetic field strengths. 

In this case, small-scale, tangled fields and the reflection of electrons from localized areas of regions of relatively strong field case can reduce the overall conductivity by a significant factor \citep{1998PhRvL..80.3077C,2001ApJ...562L.129N,2016MNRAS.460..467K}. At the same time, large-scale fields will lead to conduction that is highly anisotropic and constrained to travel in the direction of the overall field.

In \citep{2020ApJ...892...59C} (hereafter Paper III), we examined the impact of strong ($\beta=1$) magnetic fields on hot-wind cold cloud interactions and found that parallel fields have only a small effect on the lifetime of the clouds, while perpendicular fields squeeze the cloud, increasing mass loss. However, these simulations did not include the impact of electron thermal conduction on its evolution. Recently, \cite{2021MNRAS.502.1263K} studied anisotropic conduction in the interaction of cold clouds and slow winds ($\chi_0=200$ and $v_h=75$ km/s), but did not address the higher velocities that are observed on galaxy scales. Other related and recent work includes simulations by \cite{2020MNRAS.492.1841L, 2022arXiv221009722J}, which we revisit in \S5.

Here we conduct the first study of the impact of magnetic fields and thermal conduction on the evolution of radiatively-cooled clouds in the conditions that occur in outflowing galaxies. Adopting the same cloud and wind parameters as in the previous papers in this series, we explore the effect of including a heat-conduction diffusion coefficient of 10\% of the Spitzer value, to approximate the effect of small-scale magnetic fields, and explicitly including a weak large-scale magnetic field that determines the overall directionality of the electron conduction. By varying the direction of this field with respect to the wind, along with the overall parameters of the simulations, we draw general conclusions as to the effect of magnetic fields in regulating conduction in multiphase galaxy outflows.

The structure of this paper is as follows: In \S2 we review the physics of cold clouds embedded in hot plasmas. In \S3 we describe our methods. In \S4 we present our results, which we discuss and conclude in \S5.

\section{The Physics of Cold Clouds}

As described in Paper I, the ratio of the cooling time to the cloud crushing time is
\be
\frac{t_{\rm cool}}{t_{\rm cc}} = \frac{3 n_{\rm c} k T_{\rm ps} v_h}{ 2 \Lambda(T_{\rm ps}) n_{\rm e,c} \chi_0^{1/2}} \frac{1}{R_{\rm c} n_{\rm i,c} } = \frac{N_{\rm cool}}{R_{\rm c} n_{\rm i,c}  },
\ee
where $k$ is the Boltzmann constant, $T_{\rm ps}$ is the post-shock temperature (which is different from the temperature of the hot wind), $\Lambda(T_{\rm ps})$ is the equilibrium cooling function evaluated at $T_{\rm ps}$, $n_c$, $n_{\rm e,c},$ and $n_{\rm i ,c},$  are the total, electron, and ion number densities within the cloud, respectively, and $N_{\rm cool} \equiv \frac{3 n_c k T_{\rm ps} v_{\rm h} }{ 2 \Lambda(T_{\rm ps}) \chi_0^{1/2} n_{\rm e,c}}$ is a column density that depends on the velocity of the transmitted shock. This means that, if we think in units of the cloud crushing time, the evolution of a radiative cloud will depend only on the column density $R_{\rm c} n_{\rm i,c}$ rather than on size and density independently. 

As shown in Paper II, the evolution of the cloud continues to depend only on the column density even in the presence of conduction.
In the isotropic case, the classic conductive flux $\mathbf{q}_\mathrm{iso,cl}$ is given by
\be
\mathbf{q}_\mathrm{iso,cl} = \kappa(T) \nabla T,
\label{eq:thermalcond-iso-cl}
\ee
where $\kappa(T) = f_{\rm cond} \times 5.6 \times 10^{-7} T^{5/2}$ erg s$^{-1}$ K$^{-1}$ cm$^{-1}$ 
and $T$ is temperature, and the saturated conductive flux $\mathbf{q}_\mathrm{iso,sat}$
is given by
\be
\mathbf{q}_\mathrm{iso,sat} = 0.34 n_e k_{\rm B} T c_{\rm s,e} \frac{\nabla T}{|\nabla T|},
\label{eq:thermalcond-iso-sat}
\ee
where $c_{\rm s,e} = (k_{\rm B} T/m_{\rm e})^{1/2}$ is the isothermal sound speed of the
electrons in the exterior gas and $m_e$ the mass of the electron~\citep{1977ApJ...211..135C}.
The mean free path of the electrons is
$\lambda_{\rm i} = 1.3 \times 10^{18} {\rm cm}^{-2} T_7^2 n_{\rm i,c}^{-1},$ where $T_7$ is the temperature of the hot medium in units of $10^7$ K.

In the anisotropic case, the conductive flux is limited to the direction of the magnetic field~$\bhat$, such that $\mathbf{q}_\mathrm{aniso}$ is given by
\begin{align}
\mathbf{q}_\mathrm{aniso,cl} &= \kappa(T) \bhat \left( \bhat \cdot \nabla T \right), 
\label{eq:thermalanisosat}\\
\mathbf{q}_\mathrm{aniso,sat} &=
0.34 n_e k_{\rm B} T c_{\rm s,e} \mathrm{sgn} \left( \bhat \cdot \nabla T \right) \bhat .
\label{eq:thermalaniso}
\end{align}
Like cooling, this mean free path introduces a dependence on the column depth of the cloud.
Provided the column density of the cloud is less than $\lambda_{\rm i} n_{\rm i,c} = 1.3 \times 10^{18} {\rm cm}^{-2} T_7^2$ , the conducted energy will flow to the entire volume of the cloud. On the other hand, if the column depth of the cloud is greater than $\lambda_{\rm i} n_{\rm i,c}$ it will be deposited in a shell of volume $\approx 4 \pi R_{\rm c}^2 \lambda_i,$.  Inside the volume impacted by conduction, we can then compare the cooling rate to the heating rate.
For high column depth clouds with $n_{\rm i,c} R_{\rm c}> 1.3 \times 10^{18} {\rm cm}^{-2} T_7^2$, the ratio of these rates is given by
 \be
\frac{ \dot e_{\rm heat}}{\dot e_{\rm cool}}= 
\frac{4 \pi R_{\rm c}^2 \,0.34 c_{\rm s,e} n_{\rm e} k_B \, T} 
{4 \pi R_{\rm c}^2  \Lambda n_{e,c} n_{\rm i,c}  \lambda_{\rm i}}
\approx \frac{0.44}{\Lambda_{-21}} \left ( \frac{\chi_0}{1000}\right )^{-1} T_7^{-1/2}, 
\ee
assuming initial pressure equilibrium of the cloud with the exterior medium, $\chi_0 = 1000 \, T_7$ and $\Lambda_{-21}$ is the cooling rate in units of $10^{-21}$ erg cm$^3$ s$^{-1}$. Now we see that if the exterior medium is relatively hot, the stark density contrast will always guarantee that cooling is faster than conduction within the conducting region. 

However, for low column depths with $n_{\rm i,c} R_{\rm c} < 1.3 \times 10^{18} {\rm cm}^{-2} T_7^2$, 
\begin{eqnarray}
\frac{ \dot e_{\rm heat}}{\dot e_{\rm cool}}&=& 
\frac{4 \pi R_{\rm c}^2 \,0.34 c_{\rm s,e} n_{\rm e} k_B \, T} 
{(4/3) \pi R_{\rm c}^3  \Lambda n_{e,c} n_{\rm i,c} }\\
&\approx& \frac{0.44}{\Lambda_{-21} } \left ( \frac{\chi_0}{1000}\right )^{-1} T_7^{3/2}
\left ( \frac{4 \times 10^{18} {\rm cm}^{-2} }{n_{\rm i,c} R_{\rm c}}\right ),
\label{eq:heatcool}
\end{eqnarray}
which means that the heating over cooling ratio is higher the smaller the column density. As shown in Paper II, in this limit the cloud is rapidly disrupted. However, the parameters of the simulations described below are always such that $R_{\rm c} n_{\rm i,c}$ exceeds both $N_{\rm cool}$ and $n_{\rm i,c} R_{\rm c} > 1.3 \times 10^{18} {\rm cm}^{-2} T_7^2,$ such that cooling is more rapid than shock and more rapid than heating by conduction in all but the outer layers of the cloud.

Finally, while the detailed effect of small-scale tangled magnetic fields on thermal conduction in cosmic plasmas is unclear.  In particular
most theoretical estimates find that its value should lie at around 10\% of the classical Spitzer value, albeit with a large uncertainty.
As the Larmor radius of the ions in the system we are modeling is orders of magnitude smaller than the collisional mean free path, the plasma is weakly collisional, meaning that the magnetic moment of the particles $\mu =  {v_\perp}^2 /(2B)$ is conserved, where ${v_\perp}$ is the component of the particle velocity perpendicular to the magnetic field. This causes small-scale magnetic-field strength changes to be correlated with small-scale changes in the perpendicular pressure, which in turn can lead to instabilities depending on the overall plasma $\beta$.  When the degree of the pressure anisotropy is larger than $1/\beta$ this can give rise to the mirror instability, which produces small-scale fluctuations of the magnetic field that inhibit electron transport \citep{1998PhRvL..80.3077C,2016MNRAS.460..467K}.  

The result is a reduction of the effective mean free path of the electrons by a factor of $\approx 5$, which can be approximated by reducing the conduction coefficient in eq.\ (\ref{eq:thermalanisosat}) while leaving the saturated case (eq.\ \ref{eq:thermalaniso}) unchanged.  Here we assume a suppression factor of $f_{\rm cond} = 0.1,$ which to account for the possibility of additional small-scale fluctuations in the field.  This also has the added benefit of reducing the computing time by a factor of 10 over the $f=1$ case. As customary, though not necessarily correct, we also assume that the electrons and ions have the same temperature.

\section{Methods}

\subsection{Numerical Methods}t

The full system of MHD equations with radiative cooling and conduction solved in our simulations is
\begin{equation}
\partial_t \rho + \nabla \cdot (\rho \bm{u}) = 0,
\end{equation}
\begin{equation}
\rho [\partial_t \bm{u} + (\bm{u} \cdot \nabla) \bm{u} ] + \nabla \left(p + \frac{1}{2} |\bm{B}|^2 \right) - (\bm{B} \cdot \nabla) \bm{B} = 0,
\end{equation}
\begin{eqnarray}
\partial_t E &+& \nabla \cdot \left[ \left(E + p + \frac{1}{2} |\bm{B}|^2 \right)\bm{u} - \left(\bm{B} \cdot \bm{u} \right)\bm{B} \right] + \\
& & n^2 \Lambda(T) - \nabla \cdot \bm{q} = 0,
\end{eqnarray}
\begin{equation}
\partial_t \bm{B} - \nabla \times (\bm{B} \times \bm{u}) = 0,
\end{equation}
\begin{equation}
\nabla \cdot \bm{B} = 0,
\end{equation}
where $\rho$ is the density, $\bm{u}$ is the velocity, $p = k_B T \rho/(\mu m_p)$ is the pressure, $E = p/(\gamma - 1) + \frac{1}{2} \rho |\bm{u}|^2 + \frac{1}{2} |\bm{B}|^2$ is the total energy density,  and ${\bf q}$ is the electron thermal conduction.

Unlike in our previous papers, the simulations here were conducted using the performance
portable, open-source \athenapk/ code\footnote{
\athenapk/ is available at \url{https://github.com/parthenon-hpc-lab/athenapk} and commit
\texttt{f80dab9c} was used in this work.}.
\athenapk/ implements various finite volume hydro- and magnetohydrodynamics algorithms on top of  \parthenon~\citep{parthenon}, which itself is a performance-portable AMR framework
based on \athena~\citep{2020ApJS..249....4S}, \kathena~\citep{kathena}, and \kokkos~\citep{Edwards2014,Trott2021}. \athenapk/ allows for efficient simulations with AMR on various devices including GPUs from different vendors and has
demonstrated scalability with $>90\%$ parallel efficiency on up to 73728 GPUs.

Here, we use an unsplit second-order, two-stage predictor-corrector time integrator,
piece-wise parabolic reconstruction of fluxes at cell boundaries (PPM),
GLM divergence cleaning~\citep{Dedner2002}\footnote{Note that we follow the GLM implementation of \citet{Mignone2010}, i.e., the divergence wave speed is limited by the hyperbolic timestep and we use a ratio of diffusive and advective time scales of the
divergence error $\alpha = 0.4$ for the $M = 3.54$, $\chi_0 = 1,000$ simulation and 
$\alpha = 0.1$ otherwise.}, and an HLL Riemann solver. The HLL family of Riemann solvers goes back to the work by \citet{HLL} and has the distinct advantage of being computationally simple but it only captures the fast magnetosonic waves.
For convergence studies, we have also experimented with a piece-wise linear (PLM) reconstruction scheme which turned out to show a lower spectral bandwidth and smoother features (see Table 1 for an overview of runs).

In order to cope with the extreme conditions in the simulations, we implemented a first-order flux correction scheme. For cells in which the expected multi-dimensional update from the scheme above results in a negative temperature or density, all fluxes are recalculated using a forward Euler step with first-order reconstruction and LLF Riemann solver. This guarantees a positive, physically consistent update without the need to add artificial floors in the simulation.

We use the adaptive-mesh refinement capabilities of \athenapk/ to refine on the massless scalar where its concentration is $\geq 0.01$ and derefine if the concentration in the entire meshblock falls $\leq 0.001$. The size of computational box is $3000\times 1500 \times 1500$ pc$^3$ resolved with $512\times 256\times 256$ cells split into meshblock of $32^3$ cells.
It is essential that the box is large enough that it does not affect the flow around the cloud and that material does not escape laterally out of the box. The choice of our box ensures that this does not happen.
In production runs, we used two levels of refinement, yielding an effective resolution of 1500 pc/256/4= 1.46 pc, which corresponds to about 68 cells per cloud radius. In order to reduce computational costs, the adaptive refinement is disabled after 1 $\tcc$. This only pertains to the adaptive machinery so that the then static refinement, particularly around the cloud, remains in place. Finally, we use an adiabatic equation of state with a ratio of specific heats $\gamma=5/3$ and tabulated cooling using the cooling table of \citet{Schure2009} for solar metallicity.

An important length scale to keep in mind is the Field length, $\lambda_{\rm F}=\sqrt{(\kappa(T) T/(n^2 \Lambda)}$ \citep{1965ApJ...142..531F,1990ApJ...358..392M,1990ApJ...358..375B}, the maximum length scale across which thermal conduction can dominate over radiative cooling in the unsaturated case.  For 0.1 Spitzer conduction  $\lambda_{\rm F} \approx 13 \, {\rm pc} T_7^{7/4} n^{-1} \Lambda_{-22}^{-1/2}$ cm.  A second characteristic length-scale  is the cooling length $\lambda_{\rm cool}  = c_s  t_{\rm cool} \approx 2 \times 10^{13} T_7 n^{-1} \Lambda_{-22}$ cm. As described below in all our runs, $R_c = 3.08 \times 10^{20} {\rm cm}$  $n = 1$ cm$^{-3}$ within the cloud and $n = 0.001$ cm$^{-3}$ in the exterior medium.  This means that in the ambient medium $\Delta x \ll \lambda_{\rm F}, \lambda_{\rm cool}$, and inside the cloud our setup does not allow any cooling because at temperatures $T<10^4$ K, cooling is switched off.  Finally, as soon as cloud material evaporates, the resulting high pressure will not allow for fragmentation due to cooling, and thus convergence is not likely to be an issue in our simulations.

New to our simulations is the ability to include anisotropic thermal conduction. Explicit integration of diffusive physics is usually hampered by a very restrictive timestep limit that is inversely proportional to the square of the spatial resolution. When the diffusive terms are relatively large, or at very high resolution, this timestep limit can prohibit simulations for sufficient long times. Therefore, a Runge-Kutta-Legendre (RKL) super-time-stepping (STS) module \cite{meyer14} has been implemented. When STS is enabled, the diffusive equations are integrated forward in time by a separate super-time-step in an operator-split update. Each super-time-step is comprised of $s$ stages and is equivalent to $O(s^2)$ times the explicit diffusive timestep. Two operator-split super-time-steps are required in a single (M)HD update for the second-order accurate RKL2 scheme.

Finally, to account for the different nature of classic and saturated fluxes (parabolic and hyperbolic, respectively), we follow \citet{Mignone2011} and use a smooth transition
\begin{align}
 \mathbf{q} = \frac{q_\mathrm{sat}}{q_\mathrm{sat} + q_\mathrm{cl}} \mathbf{q}_\mathrm{cl},
 \label{eq:flux-transition}
\end{align}
where $q_\mathrm{sat}$ is given by Eq.~\ref{eq:thermalcond-iso-sat} and Eq.~\ref{eq:thermalaniso}.
We also employ limiters for calculating the temperature gradients following \citet{Sharma2007}.
This prevents unphysical conduction against the gradient, which may be introduced because the off-axis gradients are not centered on the interfaces.

\subsection{Simulation Suite}

As in previous papers in this series, we have carried out a suite of simulations that focuses on the conditions expected in galaxy outflows. The combination of supersonic flows, steep gradients in temperature and density, and the inclusion of radiative cooling, magnetic fields, and thermal conduction poses a formidable challenge to computational codes. 
Given the computational costs of carrying out such calculations, we have chosen only a limited set of parameters that are listed in Table \ref{tab:runs}. These runs were chosen to focus on a Mach number of 3.54, which is consistent with a cloud being impacted by a hot flow just outside of the driving radius of the galaxy, and a Mach 1 case to contrast these runs with a transonic case. Both parameter choices are also directly comparable to our results in previous papers of this series. 

As before, we simulate a spherical cloud of radius 100~pc and temperature $T_{\rm c}=10^4$~K in pressure equilibrium with the ambient gas. The cloud has a fiducial mean density of $\rho_{\rm c} = 10^{-24}$ g cm$^{-3}$, such that $R_{\rm c} n_{\rm i,c} = 1.5 \times 10^{20}$ cm$^{-2}.$
Similar to \citet{Groennow2018,2021MNRAS.502.1263K}, and unlike Papers I-IV, we use a density profile that is smooth at the cloud edges and prescribed by
\begin{align}
\rho(r) = \rho_{\rm h} + \frac{1}{2} \left(\rho_{\rm c} - \rho_{\rm h} \right) \times 
\left\{ 1 - \tanh \left[ s \left( \frac{r}{R_{\rm c}} - 1\right) \right] \right\},
\label{eq:boundary}
\end{align}
with steepness parameter $s=10$. The Jeans length in the cloud is $c_{\rm s}(G\rho)^{-1/2}\approx$ 2 kpc, which means that the cloud is pressure confined rather than gravitationally bound, and which justifies the neglect of gravity in our simulations. In order to mark the cloud material, we also include a massless scalar that is set to 1 in the initial cloud volume and set to 0 elsewhere.

\begin{table*}[tp]
\footnotesize
\caption{Parameters of our simulations. Columns show the Mach number, velocity (in units of km s$^{-1}$), plasma $\beta$, magnetic field orientation, $f_{\rm cond},$ the type of conduction and the ratio $t_{\rm cool}/t_{\rm cc}$. In all runs, the initial density contrast is $\chi_0$ = 1,000, the ambient temperature is $10^7$K corresponding to an ambient sound speed of 480 km s$^{-1},$ and the initial column density of the cloud is $R_{\rm c} n_{\rm i,c}, = 1.5 \times 10^{20}$ cm$^{-2}.$}
\begin{center}
\begin{tabular}{|l|c|c|c|c|c|c|c|c|}
\hline
\qquad Name & $M$ & $v_{\rm h}$ & $\beta$ & Orientation & $f_{\rm cond}$ & Conduction Type & Reconstruction & $t_{\rm cool}/t_{\rm cc}$\\
\hline
M3.5-hydro & 3.54  &  1700 & $\infty$ & - & 0 & - & PPM & 32\\
M3.5-par & 3.54  &  1700 & 100 & Parallel & 0 & - & PPM & 32\\
M3.5-par-iso & 3.54  &  1700 & 100 & Parallel & 0.1 & Isotropic& PPM& 32\\
M3.5-par-aniso & 3.54  &  1700 & 100 & Parallel & 0.1 & Anisotropic& PPM & 32\\
M3.5-trans & 3.54  &  1700 & 100 & Perpendicular & 0 & - & PPM& 32\\
M3.5-trans-iso & 3.54  &  1700 & 100 & Perpendicular & 0.1 & Isotropic & PPM & 32\\
M3.5-trans-aniso & 3.54  &  1700 & 100 & Perpendicular & 0.1 & Anisotropic & PPM & 32\\
M1-hydro & 1.00  &  480  & $\infty$ & - & 0 & - & PPM & 1.2\\
M1-par-aniso & 1.00  &  480  & 100 & Parallel & 0.1 & Anisotropic & PPM & 1.2\\
M1-trans-aniso & 1.00  &  480  & 100 & Perpendicular & 0.1 & Anisotropic & PPM & 1.2\\
M3.5-trans-10-PPM & 3.54  &  1700 & 10 & Perpendicular & 0 & - & PPM & 32\\
M3.5-trans-10-PLM & 3.54  &  1700 & 10 & Perpendicular & 0 & - & PLM & 32\\
M3.5-trans-10-nocool & 3.54  &  1700 & 10 & Perpendicular & 0 & - & PPM (low res) & 32\\
M3.5-trans-1-nocool & 3.54  &  1700 & 1 & Perpendicular & 0 & - & PPM (low res) & 32\\
\hline 
\end{tabular}
\end{center}
\label{tab:runs}
\end{table*}

The initial gas velocity is set to $0$ within $1.3 R_{\rm c}$ and to a fixed speed, $v_{\rm h}$, elsewhere. The gas flows in from the lower $y$-boundary with fixed $v_{\rm h}$, density and pressure, and leaves the computational domain on the upper $y$-boundary. This means that we have an inflow boundary condition on the lower $y$-boundary and outflow boundary conditions everywhere else. The magnetic fields are initialized either parallel to the flow (called parallel fields) or perpendicular to the flow (called perpendicular fields). We chose only to simulate weak magnetic fields as we are interested in the effect of the field direction on the anisotropy of thermal conduction and not on the dynamical effects of magnetic fields, although we also conduct two runs with a significant field strength to compare against our previous results, as discussed in more detail below. Hence, we set the magnetic field strength such that the ratio of the thermal pressure to the magnetic pressure, $\beta$, is 100 for the majority of our runs and $\beta = 10$ for our comparison runs. 
The full set of run parameters spanned by our simulations is described in Table \ref{tab:runs}.

\section{Results}

\subsection{Mass Evolution}

\begin{figure*}[t]
\centering
\includegraphics[width=.98\textwidth]{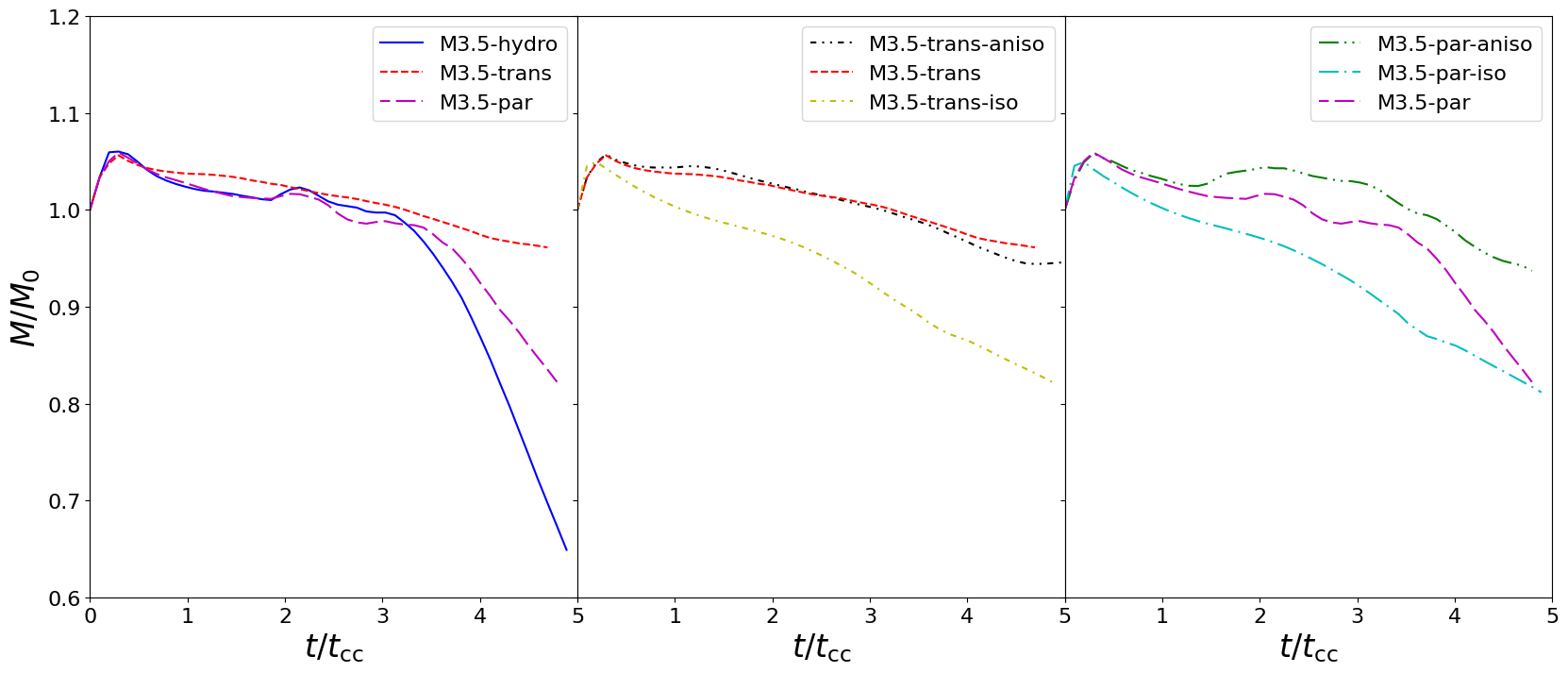}
\caption{Evolution in the mass fraction of material with $\rho \geq \rho_{\rm c}/3$ in Mach 3.5 simulations. {\em Left:} In simulations without conduction, perpendicular magnetic fields are efficient at preserving the cloud while parallel fields have little effect on the mass evolution. {\em Center:} In simulations with perpendicular magnetic fields and various types of conduction isotropic thermal conduction leads to significant mass loss that is not seen in the anisotropic conduction case. {\em Right:} In simulations with parallel fields, the isotropic thermal conduction case evolves similarly to the isotropic case with perpendicular fields, while anisotropic conduction helps preserve the cloud.}
\label{fig:massplot}
\end{figure*}

\begin{figure*}[t]
\centering
\includegraphics[width=.95\textwidth]{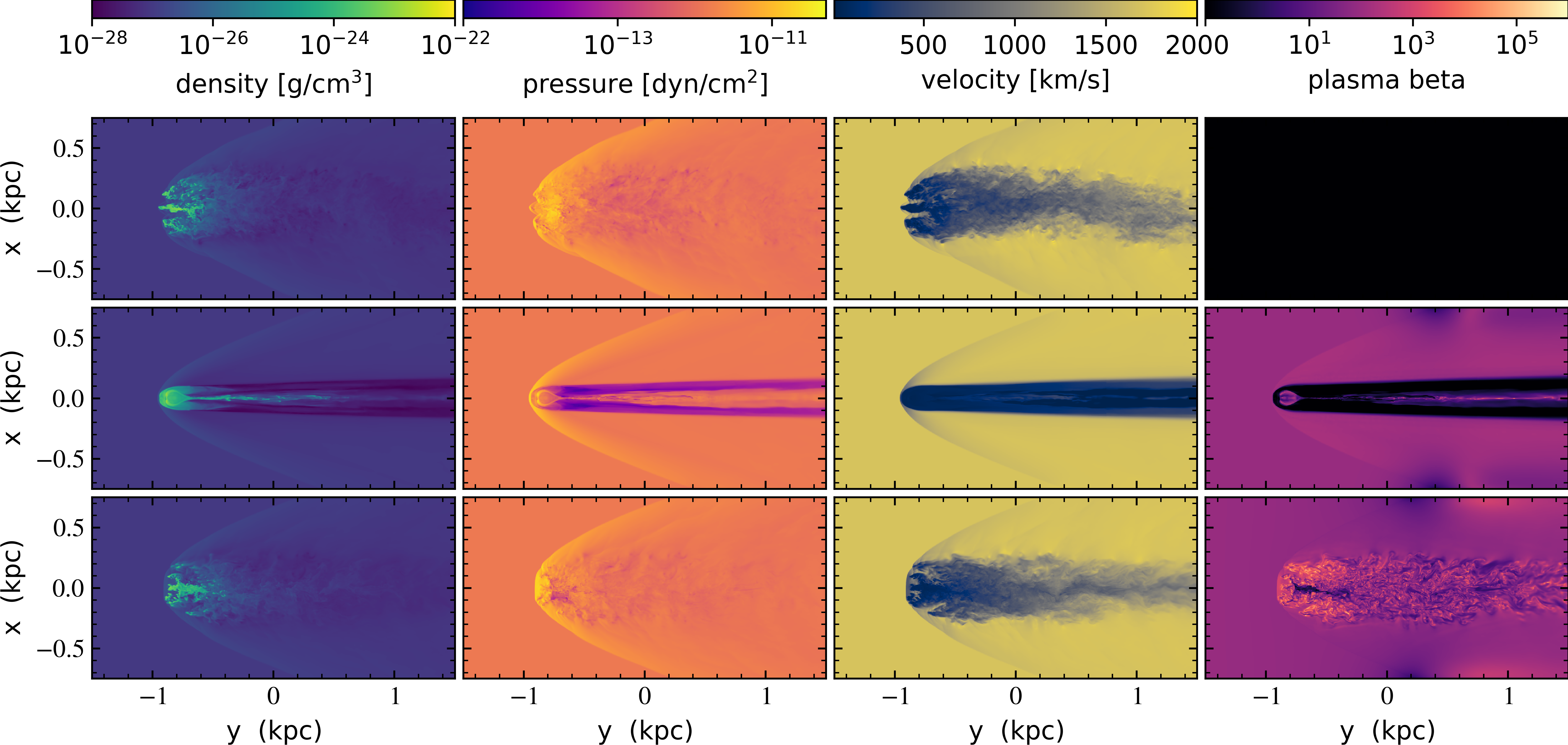}
\caption{Images of density (first column), thermal pressure (second column), streamwise velocity (third column), and plasma-$\beta$ (fourth column) at $t=4 \tcc$ from the central slice
in simulations without thermal conduction. The hydrodynamic run (top row), displays a similar overall morphology as the M3.5-par run with magnetic fields initially parallel to the flow (bottom row). However, the M3.5-trans case with initially-perpendicular magnetic fields (center row) contains a cloud that is more compact both in both directions, as well as a large magnetically-dominated region in the tail.}
\label{fig:slices_nocond}
\end{figure*}

Figure \ref{fig:massplot} shows the evolution of the dense cloud mass in our Mach 3.5 simulations. Here, as we have done in previous papers, we define cloud material as anything with a density of greater than 1/3 the initial density of the cloud. In the left plot of this figure, we contrast the evolution of the hydrodynamic case with two MHD cases with weak ($\beta=100$) magnetic fields: one in which the field is initially parallel to the direction of the winds, and one in which the field is initially aligned in the perpendicular direction.  Images of density, thermal pressure, velocity, and plasma-$\beta$ from these simulations at $t=4 \tcc$ are given in Figure \ref{fig:slices_nocond}.

In hydrodynamic interactions, the disruption of the cloud is caused by the impact of the exterior wind both along the sides of the cloud and in the direction of the flow itself \citep[e.g.][Paper I]{1994ApJ...420..213K,1995ApJ...454..172X,2004ApJ...604...74F,2005A&A...443..495M,2009ApJ...703..330C,2010MNRAS.404.1464M,2011MNRAS.415.1534M,2019MNRAS.486.4526B}. 
Along the sides of the cloud, the velocity shear leads to a mixing layer which grows at a rate $\Delta v_{\rm KH} \propto \Delta v \chi^{-1/2},$ where $\Delta v$ is the velocity difference between the cloud and the hot medium \citep{1961hhs..book.....C}, meaning that the cloud will be disrupted on a timescale $R/\Delta v_{\rm KH} \propto t_{\rm cc}.$ This rate can be understood in terms of entrainment into a spatially-growing shear layer made up of large-scale vortical structures convecting at a velocity $v_c$ \citep{1981InES....4..111C,1986AIAAJ..24.1791D}, which exists even in the presence of cooling.  However, the growth of the KH instability is slowed significantly if the exterior flow is supersonic \citep[e.g.][]{1986PhFl...29.1345C,1988JFM...197..453P, 1994JFM...259...47B, 2000JFM...414...35S} leading to a large increase in the cloud lifetimes in high-Mach number interactions \citep[e.g. Paper I,][]{2017ApJ...834..144S}.

In the streamwise direction, the most important effects are the buoyancy-driven Raleigh-Taylor (RT) instability \cite[e.g.][]{1961hhs..book.....C}, which operates effectively in the subsonic case, and the stretching of the cloud by pressure gradients, which operates effectively in the supersonic case (Paper I). If the exterior flow is subsonic, the acceleration of the dense cloud by the lower density wind leads to the growth of the RT instability at a rate $h = \alpha_{\rm b} A g t^2,$ where $g$ is the acceleration, $A = (\rho_{\rm c}-\rho_{\rm h})/(\rho_{\rm c}-\rho_{\rm h}) \approx 1$ is the Atwood number of the interaction, and $\alpha_{\rm b} \approx 0.03-0.07$ is a constant that depends weakly on the initial geometry \citep{1984PhyD...12...45R,1995PhRvL..74..534A,2004PhFl...16.1668D, 2006PhFl...18h5101D, 2008ApJ...686..927S}.  Since $g \approx \rho_{\rm h} v_{\rm h}^2 \pi R^2/ (\rho_{\rm c} \frac{4 \pi}3{R^3}) \approx \chi^{-1} v_{\rm h}^2 / R$, setting $h$ equal the size of the cloud gives a disruption time $\approx \alpha_{\rm b}^{-1/2} \chi^{1/2} R / v_{\rm h}$ or a few times $t_{\rm cc}.$

In supersonic cases, such as the ones modeled here, the evolution of the cloud is dominated by the strong difference in pressure between the front and the back of the cloud (Paper I). In this case, the front of the cloud experiences an increase in pressure approaching the $1 + M^2$ enhancement expected for a normal shock. On the other hand, the pressure increase downstream from the front of the clouds is more modest, scaling approximately as $1+M$ due to the fact that this material passes through an oblique shock. As described in Paper I, this results in the stretching of the cloud on a timescale proportional to $t_{\rm cc} \sqrt{1+M},$ which serves as the dominant mechanism leading to cloud disruption in the supersonic case.

The presence of magnetic fields has a noticeable impact on these disruption processes, even at the moderate $\beta = 100$ values taken for these simulations. In the M3.5-par case in which the field lines are initially in the direction of the flow, the primary impact of the field is to stabilize the KH instability, such that the clumping of the material in the direction perpendicular to the flow is weaker than in the hydrodynamic case. Note however, that as the KH instability is already weak in these supersonic runs, the impact of magnetic fields on the mass loss in the cloud is relatively minor.

 \begin{figure*}[t]
\centering
\includegraphics[width=.95\textwidth]{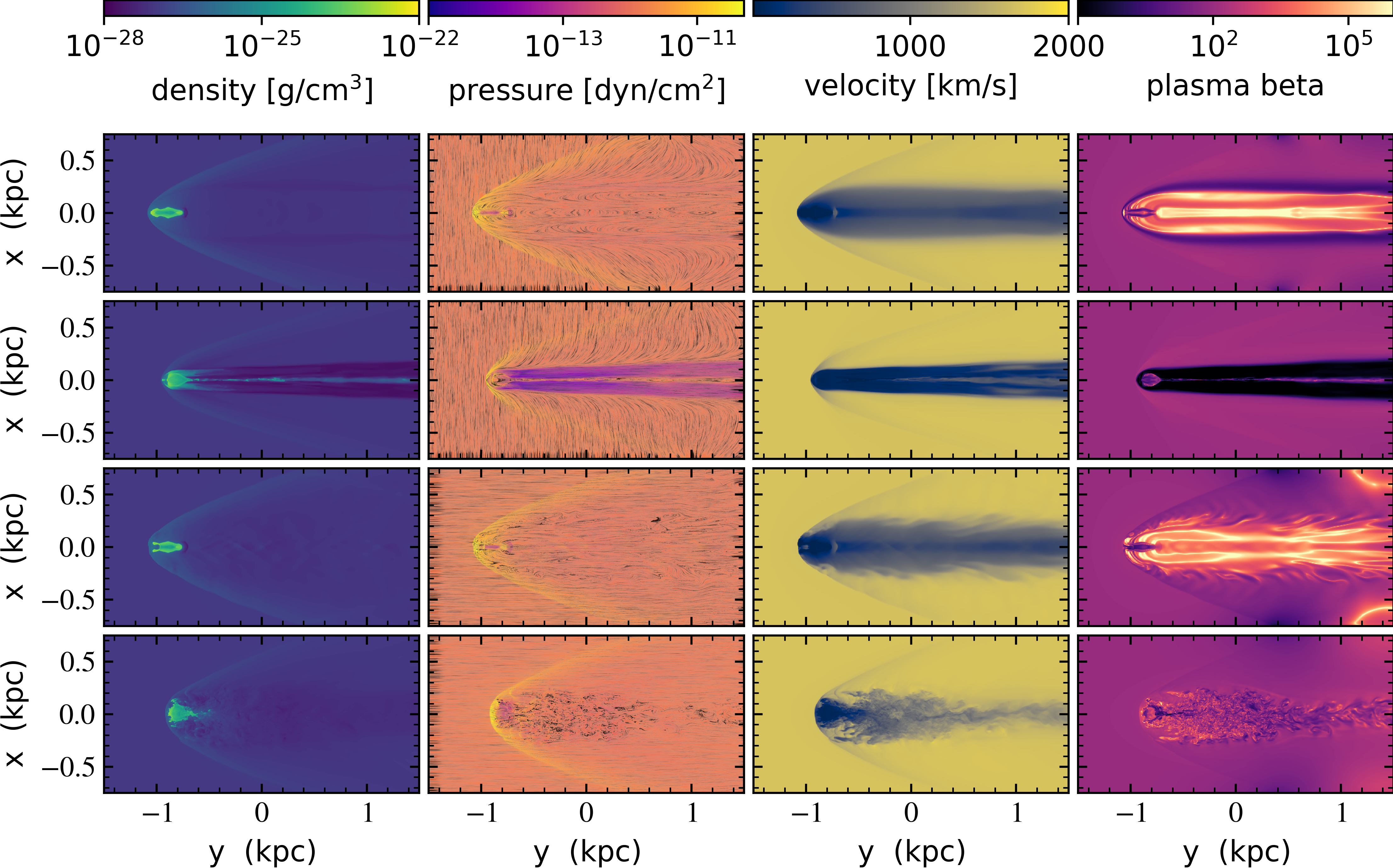}
\caption{Images of density (first column), thermal pressure and magnetic field lines (second column), streamwise velocity (third column), and plasma-$\beta$ (fourth column) at $t=4 \tcc$ from the central slice in simulations including thermal conduction. The evolution of the M3.5-trans-iso run (top row) is dominated by the evaporative flow, which compresses the cloud and isolates it from the exterior medium. On the other hand, M3.5-trans-aniso run (second row) shows little evidence of an evaporative flow, as most of the field lines traveling through the cloud are bent into the mostly-cold tail. The M3.5-par-iso run (third row) displays a morphology very similar to the isotropic conduction case with initially-perpendicular fields. Finally the M3.5-par-aniso run (fourth row) shows a weak evaporative flow that helps to preserve the cloud without causing rapid mass-loss.}
\label{fig:slices_cond}
\end{figure*}

In the case in which the field is oriented perpendicular to the flow, however, the impact of the magnetic field on the mass loss is much more significant. In the M3.5-trans simulation, the field has two major effects: first is stabilizes the RT and KH instabilities through magnetic draping \citep[e.g.][]{2008ApJ...677..993D,2015MNRASBB,2021MNRAS.502.1263K}, leading to less clumping than seen in the other two cases. Secondly, and more importantly, it acts to counteract the stretching of the cloud by the pressure gradients. This is due to the fact that stretching in the streamwise direction leads to a separation between the magnetic field lines threading the cloud, and a corresponding drop in field strength due to flux freezing. As the magnetic pressure is proportional to $B^2$ this leads to a significant drop in the pressure within the cloud, which opposes the expansion decreasing the mass loss in the cloud. Finally, the compression of magnetic fields due to the bending of field lines leads to a magnetically-dominated sheath behind the cloud, which is visible as the low-$\beta$ regions surrounding the highly collimated tail visible in the last row of Figure~\ref{fig:slices_nocond}.

The central panel of Figure \ref{fig:massplot} compares the mass evolution in three cases with perpendicular magnetic fields and different levels of conduction: no-conduction (repeating the case in the left panel), isotropic thermal conduction, and anisotropic thermal conduction. Corresponding images density, thermal pressure, field-lines, velocity, and plasma-$\beta$ from the conducting simulations at $t=4 \tcc$ are given in the top two rows of Figure~\ref{fig:slices_cond}.

 A direct comparison of the mass evolution and morphology between M3.5-trans case without conduction to the M3.5-trans-iso case with isotropic thermal conduction reveals strong differences.  As was observed in \cite{2016ApJ...822...31B} in the case with isotropic conduction, a strong evaporative wind develops, which compresses the cloud due to ram pressure \citep[See also][] {2002A&A...395L..13M,2004ApJ...604...74F,2005A&A...444..505O,2013ApJ...766...45J}. The result is a much denser, compact system, which remains largely decoupled from the outside flow, losing mass at a rate that can be understood from a balance of impinging thermal energy and evaporation. The evaporative wind is almost completely dominated by ram-pressure and thermal pressure, with magnetic fields playing only a minor role, as evidenced by the large plasma-$\beta$ observed in this region.

In strong contrast to the isotropic conduction case, the mass loss rates and morphology in the M3.5-trans-aniso run are almost identical to the run without thermal conduction. This is because the magnetic fields passing through the cloud are bent back into a magnetically-dominated sheath behind the cloud, where the thermal pressure is small. This means that there is very little thermal energy available to be carried in the direction of the field lines and into the cloud, causing conduction to be strongly reduced.

Finally, the rightmost panel of Figure \ref{fig:massplot} compares the mass evolution in three cases with initially-parallel magnetic fields and different levels of conduction. Corresponding images from the conducting simulations at $t=4 \tcc$ are given in the bottom two rows of Figure \ref{fig:slices_cond}. As in the M3.5-trans-iso case, the inclusion of isotropic thermal conduction in the M3.5-par-iso run results in a strong evaporative wind that compresses the cloud and largely decouples in from the outside flow. Furthermore, because the strength of this wind is such that the thermal and ram-pressure dominate over magnetic forces, the overall evolution of the M3.5-trans-iso and M3.5-par-iso runs are very similar. Both clouds lose mass at a roughly constant rate, set by the energy input evaporating the cloud. Here we should also comment on the bump in the mass evolution at early times. This was not observed in our papers I-IV. The difference stems from a difference in our initial conditions. In papers I-IV the cloud was modelled as a hard sphere whereas here we taper the edges of the cloud according to eq.\ (\ref{eq:boundary}) in order to avoid unphysically steep gradients. As a result, the impinging wind piles up in front of the cloud leading to an initial rise of the cloud mass. However, this is only a transient effect that does not affect our results at later times.

Finally, in the M3.5-par-aniso case, the cloud mass loss rate is even slower than the M3.5-par case without conduction. This is due to the fact that the radiative wind is largely suppressed except in small regions near the front and back of the cloud. Near the front of the cloud, electrons flow along fields lines from the upstream material powering an evaporative wind which insulates it from the development of the RT instability. At the same time, material at the back of the cloud is able to conduct heat from the wake region, which remains hot and thermally dominated in this case. This balances the cloud from expansion in the streamwise direction, also significantly extending its lifetime.

\subsection{Velocity Evolution}

\begin{figure*}[t]
\centering
\includegraphics[width=.99\textwidth]{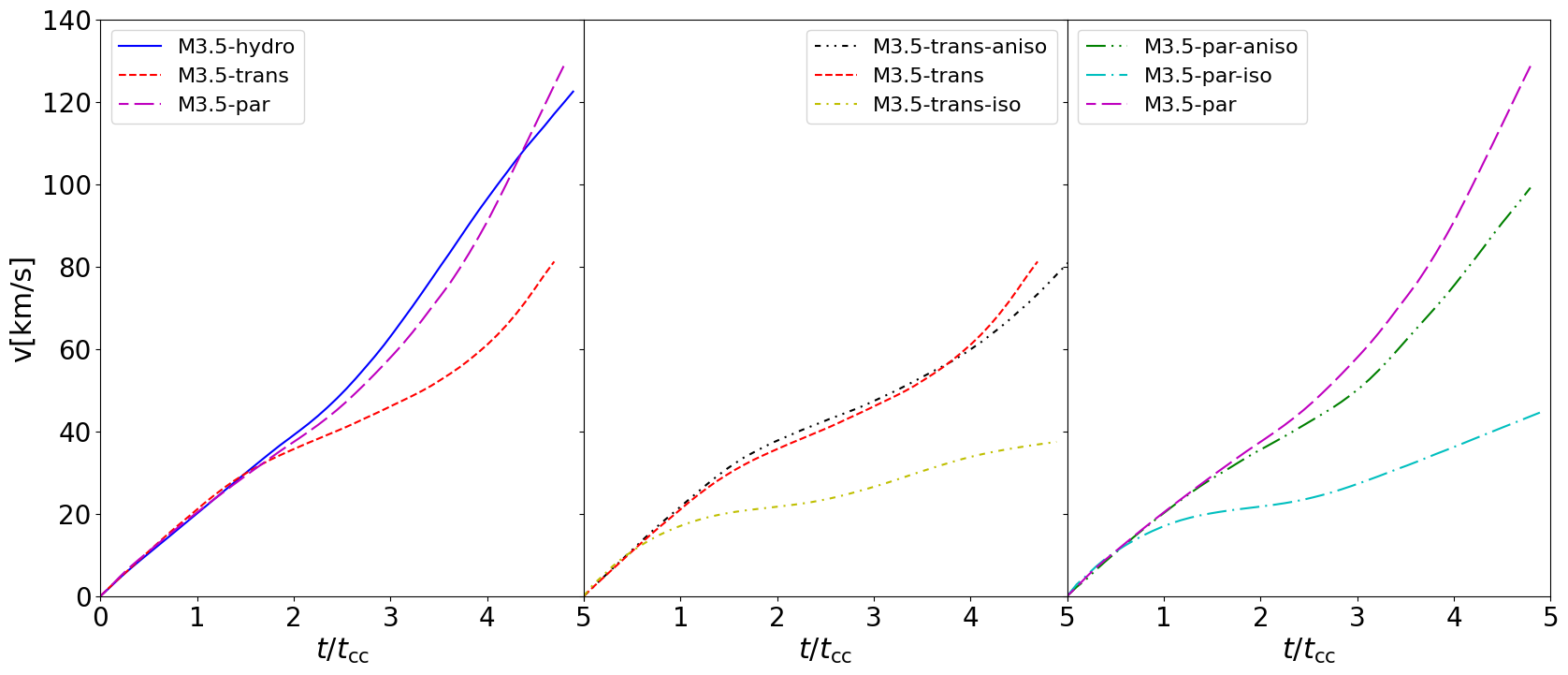}
\caption{Impact of anisotropic conduction on the evolution of the velocity of the material with $\rho>\rho_{\rm c}/3.$ As in Fig.\ \ref{fig:massplot}, the upper panels compare runs with fields parallel to the flow direction, and the lower panels compare runs with the fields perpendicular to the flow direction. The left panels compare runs without thermal conduction, the central panels compare runs with perpendicular fields, and the right panels compare runs with parallel fields.}
\label{fig:velplot}
\end{figure*}

\begin{figure*}[t]
\centering
\includegraphics[width=.99\textwidth]{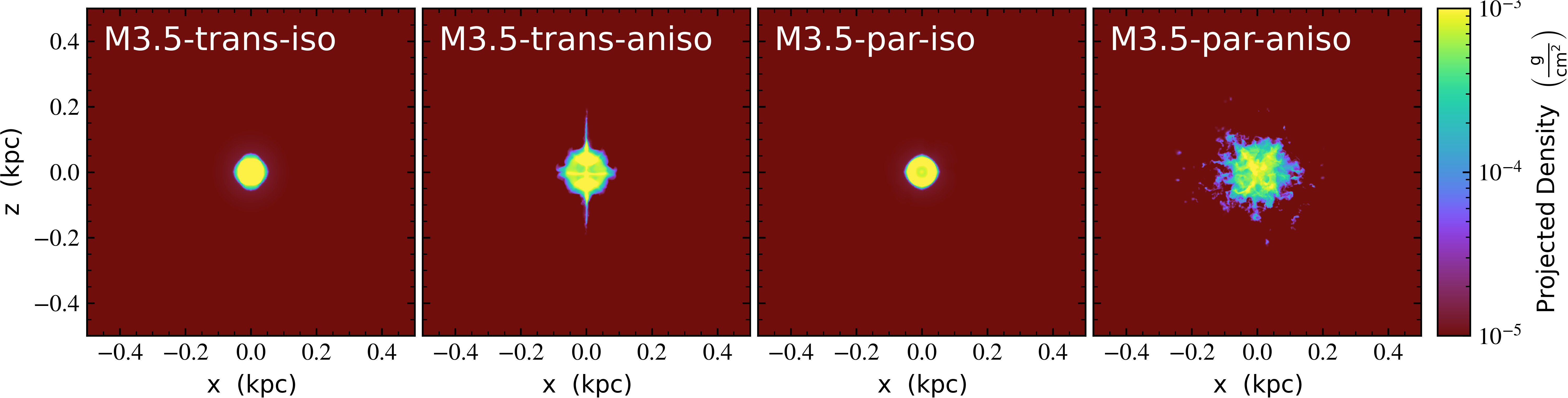}
\caption{Streamwise projection of the density of the cloud for the M3.5 simulations with thermal conduction, isotropic and anisotropic, as indicated in the panel labels. Note that the image does not show the entire width and height of the simulation box.}
\label{fig:crosssect}
\end{figure*}

Next we turn to the acceleration of the cloud by the supersonic wind.  In Fig.~\ref{fig:velplot} we show the evolution of the velocity of the material with $\rho > \rho_{\rm c}/3,$ with lines and panels matching those used in Fig.~\ref{fig:massplot}. As shown in previous work, in the hydrodynamic case, the hot wind is inefficient at accelerating the cold cloud. This also proves to be true for all these cases studied here, such that even after five cloud crushing times the cold material attains speed between 70 and 120 km/s, which is less than 10 \% that of the ambient medium. 

Comparing the various runs shows that in many ways the cloud velocities follow the same trends as the mass evolution. In cases without conduction, we find that the streamwise velocity of the cloud is highest in the purely hydrodynamic and M3.5-par runs, whereas the cloud is about 25\% slower for the M3.5-trans runs (after 5 $\tcc$).  Switching on anisotropic conduction has very little effect on these results, such that the velocity evolution in the M3.5-trans- and M3.5-par-aniso closely follow the cases without thermal conduction. 

However, as also found in \cite{2016ApJ...822...31B}, in the runs with isotropic conduction, the clouds are only accelerated to half the speed as in the anisotropic case. This is because the evaporative pressure compresses the cloud such that it presents a smaller cross-section to the impinging wind. This can be seen, for instance, in the streamwise projection of the density, which is shown in Fig.~\ref{fig:crosssect}.

Note, however, that the velocity evolution for the runs without conduction is different from the case of strong magnetic fields. In \cite{2020ApJ...892...59C}, it was shown that the clouds in $\beta=1$ perpendicular fields attained a higher velocity than the clouds in parallel fields (see fig.~12 in \cite{2020ApJ...892...59C}). Apparently, perpendicular fields lead to additional acceleration if the fields are strong. This higher acceleration was attributed to a hundred-fold increase in magnetic pressure at the front of the cloud as the perpendicular field lines are compressed by the flow. 

We confirmed this by comparing fixed grid simulations of our setup at $\beta=100$, $\beta=10$ and $\beta=1$ (not shown). These confirmed that the $\beta=1$ case is markedly different from the lower field cases, in that it leads to a significantly stronger acceleration in the perpendicular field case.

In all cases the magnetic tension exceeds the effect of the magnetic pressure in front of the cloud. When $\beta$ becomes as low as 1, the tension become so strong as to change the flow around the cloud. The opening angle of the shock around the cloud increases and the stand-off distance of the shock increases. The wind thus needs to push against a``magnetic wall'' that couples to the cloud and leads to an higher acceleration than in cases with higher $\beta$. 

\subsection{Low-Mach Number Interactions}

\begin{figure*}[t]
\centering
\includegraphics[width=.95\textwidth]{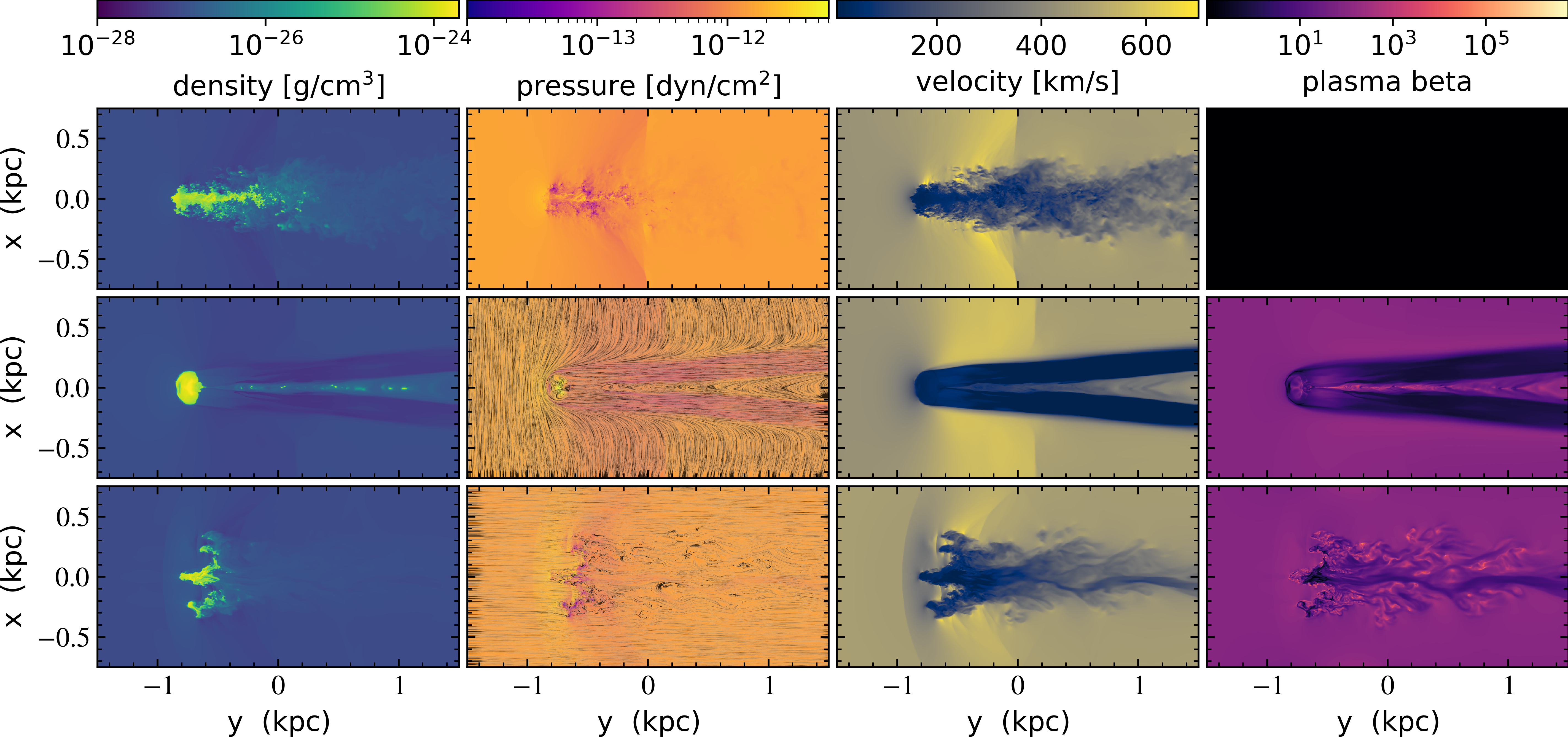}
\caption{Images of density (first column), thermal pressure and magnetic field lines (second column), streamwise velocity (third column), and plasma-$\beta$ (fourth column) at $t=3 \tcc$ from the central slice in simulations including anisotropic thermal conduction. The top row shows the hydro run, the second row the M1-trans-run, and the third row the M1-par run.}
\label{fig:slices_480}
\end{figure*}

\begin{figure*}[t]
\centering
\includegraphics[width=.8\textwidth]{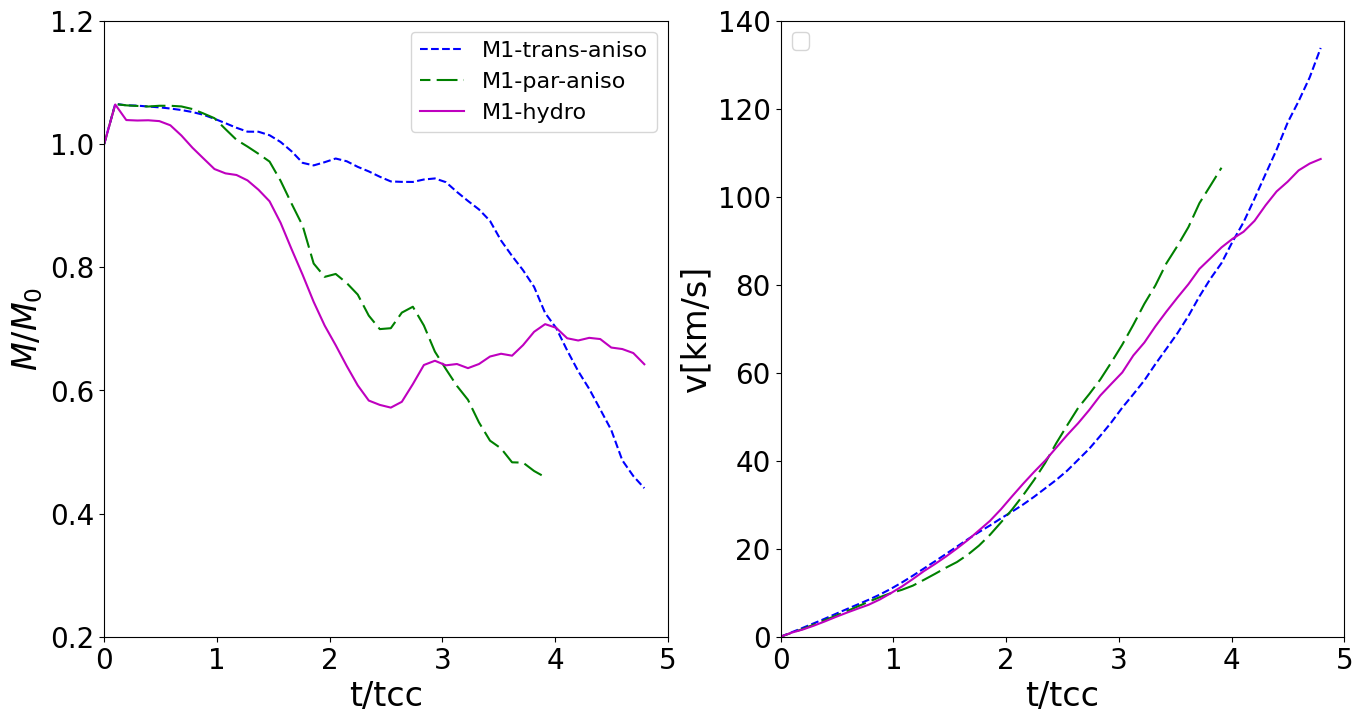}
\caption{Impact of anisotropic conduction on the evolution of the mass and velocity of the material with $\rho>\rho_{\rm c}/3$ for a $M=1$ run with $v=480$ km/s.}
\label{fig:massvelplot480}
\end{figure*}

To compare the evolution of a cloud surrounded by a supersonic medium to one impacted by a transonic wind, we repeated our simulations reducing the ambient wind speed to 480 km/s, but leaving all other parameters untouched, which results in a Mach number of $M\approx1$. Here we only looked at a single, purely hydrodynamic run (M1-hydro) and two runs with anisotropic conduction and different initial magnetic field orientations (M1-par-aniso and M1-trans-aniso). The mass and velocity evolution of these runs are shown in Fig.~\ref{fig:massvelplot480}, and corresponding slices of key parameters at $t= 3 \tcc $ are shown in Fig.~\ref{fig:slices_480}. 

As seen in previous simulations (e.g.\ Paper 1) at lower wind speeds, the pressure differences across the cloud are reduced, but the RT and KH instabilities become more efficient, leading to a more expedient mass loss. Thus, in the hydrodynamic case, we can see that by $3 t_{\rm cc}$ the cloud is highly disrupted, with much of the mass fragmented into a tail made up of a series of small clumps. Note that unlike the M3.5-hydro run, the strong KH that exists in transsonic interactions means that the density sheared material drops well below the $\rho_{\rm c}/3$ threshold that we use to define the boundary of the cloud. As a result the effective cross section of the cloud drops well below $\pi$ times the initial cloud radius squared.

In the case M1-trans-aniso with perpendicular fields and anisotropic conduction, on the other hand, draping of the magnetic fields acts to preserve the cloud from instabilities, leading to a more ordered flow and lower overall mass-loss rate. Interestingly, because of the strong KH seen in the M1-hydro run, this run's cross section of dense material is greater in the case with perpendicular fields than in the hydro case. Finally, as in the higher Mach number M3.5-trans-aniso simulation, the field lines are bent behind the cloud, forming a low-beta region in which the thermal energy density is low. Also this greatly reduces the ability of electrons to heat the cloud, suppressing the role of evaporation in the interaction (as in the M3.5-trans-aniso run).

On the other hand, in the M1-par-aniso run with anisotropic conduction and fields in the direction of the flow, electrons can flow freely from the hot upstream material. The resulting evaporative flow then plays two key roles in the overall cloud evolution. First, evaporation in the flow direction leads to the presence of a shock that precedes the cloud, reducing the velocity of the flow as it passes over the cloud itself. This reduces the growth of instabilities, preserving the cloud for longer times and reducing mass loss into the tail. However, a second effect of conduction also is to cause evaporation between the dense regions that arise from the suppressed instabilities.  This causes clumps to move away from each other even in the direction perpendicular to the field lines. Interestingly, this causes the transverse size of the cloud M1-par-aniso to be even larger than in the hydro run. However the overall mass loss is similar between the M1-par-aniso and M1-hydro runs.

A striking feature in the left panel of Fig.~\ref{fig:massvelplot480} is the behavior of the purely hydrodynamical run. After 2.5 $\tcc$ the total mass with density above $\rho_{\rm c}/3$ increases again. This is caused by condensation in the tail of the cloud as discussed in \cite{2018MNRAS.480L.111G}. They find that in clouds with radii larger than a critical radius, radiative cooling will lead to eventual mass growth. 
For a cloud that is disrupted in $\tilde t$ cloud crushing times, the temperature of the mixed gas is
\begin{equation}
T_{\rm mix}=\frac{T_{\rm c}\dot{m}_{\rm c} + T_{\rm ps}\dot{m}_{\rm w}}{\dot{m}_{\rm c} + \dot{m}_{\rm w}} \approx \tilde A \tilde t \chi^{-1/2} T_{\rm ps},
\end{equation}
where $\dot{m}_{\rm c} = (4 \pi/3) R_c^3 \rho_h \chi / (\tilde t \, t_{\rm cc})$, $\dot{m}_{\rm w} = \pi \tilde A R_c^2 \rho_h v_h$, $T_{\rm ps}$ is the post-shock temperature, and $\tilde A$ is a factor that describes the area of the impinging gas that is mixed into the layer in units of $\pi R_c^2.$  
With $\tilde A \tilde t \approx 1$, in the transonic case $\chi^{-1/2} T_{\rm ps} \approx \sqrt{T_c T_h}$ and this reduces to eq.\ (1) in \cite{2018MNRAS.480L.111G}.

This critical radius is then found by setting the cooling time in this mixing gas to $\eta$ times the cloud-crushing time. Again assuming $\tilde A \tilde t \approx 1$ this gives
\begin{equation}
R_{\rm crit} = \frac{v_h t_{\rm cool,mix}}{\chi^{1/2}} \eta^{-1} \approx 2\, {\rm pc}\,\frac{T_{\rm c,4}^{5/2}M}{P_3\Lambda_{-21.4}(T_{\rm mix})} \frac{\chi_0}{100} \eta^{-1} \frac{T_{\rm ps}^2}{T_{\rm w}^2},
\end{equation}
where $T_{\rm c,4}$ is the cloud temperature in units of $10^4$~K, $P_3=n_ hT/({\rm cm^3K})$, $\Lambda_{-21.4}$ is the cooling rate in units of $10^{-21.4}$ erg cm$^3$ s$^{-1}$ and $T_{\rm mix}$ is the temperature of the mixed gas. This reduces to eq.\ (2) in \cite{2018MNRAS.480L.111G} in the transsonic case.

Plugging in the parameters for our $M=1$ simulation, we find that our cloud radius of 100 pc is safely above $R_{\rm crit} \approx 2$ pc. Hence, the growth of the cloud mass due to the condensation is not unexpected. However, we do not find this behavior in our $M=3.5$ runs in which the postshock temperature and mixing temperature are both increased by a factor of $1+M^2$, leading to a critical radius that is $\approx 1000$ times greater.

We also do not see much condensation in the $M=1$ runs with magnetic fields since the fields suppress the mixing that leads to radiative condensation. However, there is some cold gas in a thin filament in the wake of the runs with perpendicular magnetic fields (see Fig.~\ref{fig:slices_cond}) that can be attributed to cooling. The exact critical sizes of clouds are also subject to debate \citep[see e.g.][]{2020MNRAS.492.1841L} and we are aware of work that looks at this problem for magnetized winds \citep{2023arXiv230409897H}.

Finally, the right panel of Fig.~\ref{fig:massvelplot480} shows the velocity evolution of the dense $\rho = \rho_{\rm c}/3$ material in the three cases. As was true at higher Mach numbers, the evolution of the run with parallel fields and anisotropic conduction shows a similar velocity evolution to the hydro case.  This is because while the M1-par-aniso run contains clumps that have moved away from each other due to evaporation, the overall cross-section of the clumps remains similar to the M1-hydro case. Because the cross section of dense material is greater in the M1-trans-aniso case than in the M1-hydro run, the momentum transferred to the cloud exceeds that in the other two cases. This leads to a similar velocity evolution to the hydro case, even though there is much less mass loss in the M1-trans-aniso run.

\subsection{Comparison To Previous Work}

From a technical point of view, there are multiple differences between this investigation and previous work. Most importantly, in Paper II where we studied the impact of thermal conduction in the purely hydrodynamic case, the saturation of the conductive flux was implemented by taking the minimum value of the classic and saturated flux. Here, we use a smooth transition to account for parabolic and hyperbolic nature of the fluxes, cf., eq.~\eqref{eq:flux-transition}. Moreover, thermal conduction was treated using the general implicit diffusion solver in \flash whereas we employ an explicit, second-order accurate RKL2 super-time-stepping scheme. Finally, we used different cooling tables (though both assume solar metallicity).
 
The difference in cooling tables also applies to the simulations in Paper III in which we studied the impact of strong magnetic fields \citep{2020ApJ...892...59C}. While we do not expect significant differences between the divergence cleaning approach (a purely parabolic cleaning scheme in \flash versus the mixed hyperbolic-parabolic scheme in this paper), the spatial reconstruction methods matter. In Paper II we employed a second-order accurate scheme whereas we use a third-order scheme here resulting in an extended spectral bandwidth and more small-scale structure in the tail of the cloud. We reproduced the results of Paper III by reverting to a PLM reconstruction scheme. It is interesting to note that the third-order scheme employed here leads to a higher effective resolution in the simulation and also shows much better convergence properties than the runs with PLM reconstruction. Some results of runs with differing spatial resolution are shown in the appendix.

In addition to the reconstruction scheme, we also had to be careful about the choice of the Riemann solver. In previous work, we had used the Harten-Lax-van Leer discontinuities (HLLD) approximate Riemann solver developed by \cite{2005JCoPh.208..315M}. In our setup, with supersonic flows with strong density contrasts, this shock-capturing scheme  led to artificial heating in isolated cells and numerical instability that is related to the so-called carbuncle instability. This numerical instability affects the numerical capturing of shock waves and was first noticed by \cite{peery88} while simulating a high speed flow around a blunt body. The nature of this instability is still not fully understood and is usually mitigated by adding numerical dissipation \citep[see e.g.][]{FLEISCHMANN2020109004, Minoshima2021}. We tried the LHLLD solver by \citep{Minoshima2021} but still encountered numerical instabilities in a subset of the simulations. We found that the combination of PPM reconstruction schemes with HLL Riemann solvers yielded the best results in terms of effective resolution and numerical stability though at the cost of the Riemann solver not resolving the contact discontinuity.

Our work is complementary to recent work by \cite{2022arXiv221009722J} who studied clumpy and uniform clouds in transverse magnetized, subsonic flows. They found that clumpiness has a large effect on the evolution of the cloud, but they did not include conduction. It will be interesting to study the effect of clumpiness also in our setup. It is also complementary to recent work by \cite{2023MNRAS.518.5215J} who looked at the effect of anisotropic thermal conduction on clouds in the (hotter) intracluster medium.

\section{Discussion and Conclusion}

We have explored the impact of isotropic and anisotropic thermal conduction on the evolution of radiatively-cooled, cold clouds embedded in hot magnetized winds. Using the adaptive mesh refinement code \athenapk/, we modelled cloud mass loss and acceleration in supersonic and transonic flows with magnetic fields that are initially aligned parallel and perpendicular to the flow. Our simulations adopted a heat-conduction diffusion coefficient of 10\% of the Spitzer value, to approximate the effect of small-scale magnetic fields, and explicitly include a weak ($\beta=100$) large-scale magnetic field that determines the overall directionality of the conduction. 

The presence of even weak magnetic fields has a distinct impact on the disruption of cold clouds. In the M3.5-par case where the field lines are initially in the direction of the flow, the primary impact of the field is to stabilize against the KH instability. However, since the KH instability is already weak in these supersonic runs, the impact on the mass loss is relatively minor.

On the other hand, in the case where the field is perpendicular to the flow, the impact of the magnetic field on the evolution is much stronger. First, it stabilizes the RT and KH instabilities through magnetic draping \citep[e.g.][]{2008ApJ...677..993D,2015MNRASBB,2021MNRAS.502.1263K}. Secondly, it acts to counteract the stretching of the cloud by the pressure gradients because stretching in the streamwise direction leads to a significant drop in the magnetic pressure within the cloud, opposing the expansion. Finally, the compression of magnetic fields due to the bending of field lines leads to a magnetically-dominated sheath behind the cloud.

Adding thermal conduction to our runs has an impact that depends strongly on assumptions about isotropy. In cases with isotropic thermal conduction, an evaporative wind forms, which dominates the evolution, stabilizing against instabilities and leading to a mass loss rate that matches the hydrodynamic case. On the other hand, in the anisotropic case, the impact of conduction is more limited and strongly dependent on the field orientation. 

In the anisotropic case with initially perpendicular fields, the mass loss rates and morphology are almost identical to the run without thermal conduction. This is because the magnetic fields passing through the cloud are swept back into a magnetically-dominated sheath trailing the cloud, where the thermal pressure is small, and not much thermal conduction takes place.

In the anisotropic case with initially parallel fields, however, conduction aids cloud survival by forming a radiative wind only near the front of the cloud, which suppresses instabilities with minimal mass loss. Near the front of the cloud, electrons flow along field lines from the upstream material powering an evaporative wind that insulates it from the development of the KH instability. At the same time, the material at the back of the cloud is able to conduct heat from the wake region, which remains hot and thermally dominated in this case. This balances the cloud from expansion in the streamwise direction, significantly extending its lifetime.

When we reduce the wind speed to $M\approx1$, the clouds break up earlier in the hydrodynamic case because the RT and KH instabilities become more efficient. In the transsonic run with perpendicular fields and anisotropic conduction, on the other hand, draping of the magnetic fields acts to preserve the cloud from instabilities, leading to a more ordered flow and lower overall mass-loss rate. Finally, in the transsonic run with parallel fields and anisotropic conduction, electrons flow freely from the hot upstream material,  resulting in an evaporative flow that plays two key roles in the overall cloud evolution.  First, evaporation in the flow direction leads to the presence of a shock that precedes the cloud, reducing the velocity of the flow and the resulting growth of instabilities, preserving the cloud for longer times.  Secondly, evaporation between the dense fingers that arise from the instabilities results in clumps that move away from each other in the direction perpendicular to the field lines. Interestingly, this causes the transverse size of the cloud M1-par-aniso to be even larger than in the hydro run. However, the overall mass loss is similar between the M1-par-aniso and M1-hydro runs, which in turn are not radically different from the rates for M1-trans-aniso.

At late times in the hydrodynamic case, we observe that the cloud mass grows eventually owing to the condensation of material in the wake. \cite{2020MNRAS.492.1841L} performed a large parameter study of cloud-crushing simulations including a wide range of physical effects, including anisotropic thermal conduction. Using a Lagrangian code and focussing on subsonic cases, they find that conduction can increase the survival time of the clouds. In our runs though, we find that magnetic fields are more efficient at preserving the clouds than in \cite{2020MNRAS.492.1841L} since the shock amplifies the magnetic fields to higher values around the cloud. Overall, we find that anisotropic conduction makes less of a difference than the effects of the ambient magnetic fields.

In all cases studied here, we find that the hot wind is inefficient at accelerating the cold cloud, leading to accelerations to $<10$ \% of the speed of the ambient medium (see Fig.~\ref{fig:velplot}). Even after five cloud crushing times, the cold material attains velocities between 70 and 120 km/s. The cloud velocities follow in some ways the same trends as the mass evolution. In the runs with isotropic conduction, the clouds are only accelerated to half this speed because the evaporative pressure compresses the cloud such that it presents a smaller cross-section to the impinging wind. However, anisotropic conduction behaves very differently and has very little effect on the acceleration of the cloud, leading to a velocity evolution that closely follows the case without thermal conduction.

In this work, we have assumed that the coherence length of the magnetic field is larger than the box size. Many effects discussed here require that the coherence length of the field is larger than the cloud, as was for example explored in the case of underdense bubbles in \cite{2007MNRAS.378..662R}. If the dominant fields have coherence lengths below the size of the cloud, the magnetic fields could have the opposite effect of expediting their destruction. Observationally, we do not know much about the topology of magnetic fields in outflowing galaxies. However, recent work using Faraday rotation of polarized background sources has shown that star-forming galaxies contain large-scale fields in the CGM \citep{2023A&A...670L..23H}.

While we have shown that magnetized, radiative and conductive models of cloud crushing show extended lifetimes, we still cannot fully solve the entrainment problem, i.e. the observation that cold, atomic clouds can attain high velocities galaxy outflows. However, the structure of the clouds in our simulations has interesting implications for observations of cold clouds, as it makes clear that the covering fractions and column densities of clouds will depend sensitively on the properties of ambient magnetic fields and the nature of thermal conduction.

In addition to galaxy outflows, our study is relevant for understanding multiphase flows in the circumgalactic medium, which is threaded by magnetic fields and subject to the anisotropic thermal conduction that comes with them.  Because the survival times for cold clouds that we find here do not depend heavily on field direction in the presence of anisotropic thermal conduction, we can hope to devise subgrid models for the evolution of cold clouds \cite[e.g.]{2020MNRAS.497.2586H,2022MNRAS.509.6091H} that can be included in large-scale simulations of galaxy evolution that do not include anisotropic thermal conduction or have the spatial resolution to capture these effects directly.

Finally, our work is also applicable to cool-core galaxy clusters, whose centers sometimes show filamentary emission of atomic gas ($<10^4$ K) embedded in the hot ($10^7$ K) intracluster medium. This cold gas is important for feeding the supermassive black hole in the central galaxy and regulating the feedback cycle that operates in these clusters.   As shown in \citep{2023MNRAS.518.5215J}, fully understanding these systems also will require a better understanding of the role of magnetic fields and anisotropic thermal conduction.

\begin{acknowledgements}

We thank Max Gronke, Ryan Farber, and Fernando Hidalgo-Pineda for valuable discussions, as well as the anonymous referee for a very constructive report.
MB acknowledges support from the Deutsche Forschungsgemeinschaft under Germany's Excellence Strategy - EXC 2121 ``Quantum Universe" - 390833306 and the support and collaboration with the Center for Data and Computing in Natural Sciences (CDCS). ES acknowledges support from the NASA grant 80NSSC22K1265. This project has received funding from the European Union's Horizon 2020 research and innovation programme under the Marie Sklodowska-Curie grant agreement No \texttt{101030214}. We would like to thank the Kavli Institute for Theoretical Physics and the organizers of the workshops Fundamentals of Gaseous Halos (Halo21) and The Cosmic Web: Connecting Galaxies to Cosmology at High and Low Redshift (CosmicWeb23). This research was supported in part by the National Science Foundation under Grant No.\ NSF PHY-1748958. Part of the simulations were run on the JUWELS supercomputer at the J\"ulich Centre for Supercomputing within the project ANISOCOND.
\end{acknowledgements}

\software{\athenapk/ and \parthenon \citep{parthenon}, yt \citep{2011ApJS..192....9T}}

\bibliographystyle{apj}
\bibliography{Conduction}

\eject

\appendix

\section{Convergence}

\begin{figure}[h]
\centering
\includegraphics[width=.5\textwidth]{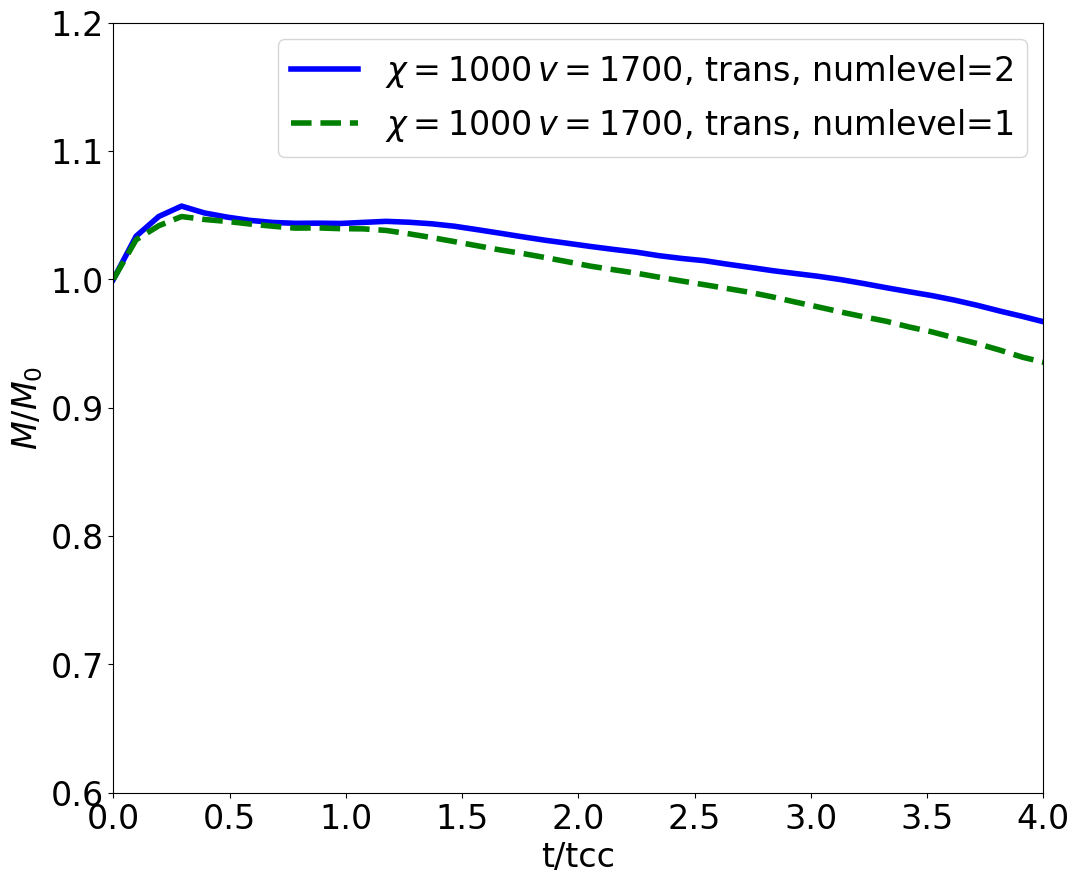}
\caption{Impact of resolution on the evolution of the mass evolution with $\rho>\rho_{\rm c}/3.$ The figure compares runs M3.5-trans-aniso with 1 and 2 levels or refinement.}
\label{fig:convergence}
\end{figure}

\begin{figure}[h] 
\centering
\includegraphics[width=.95\textwidth]{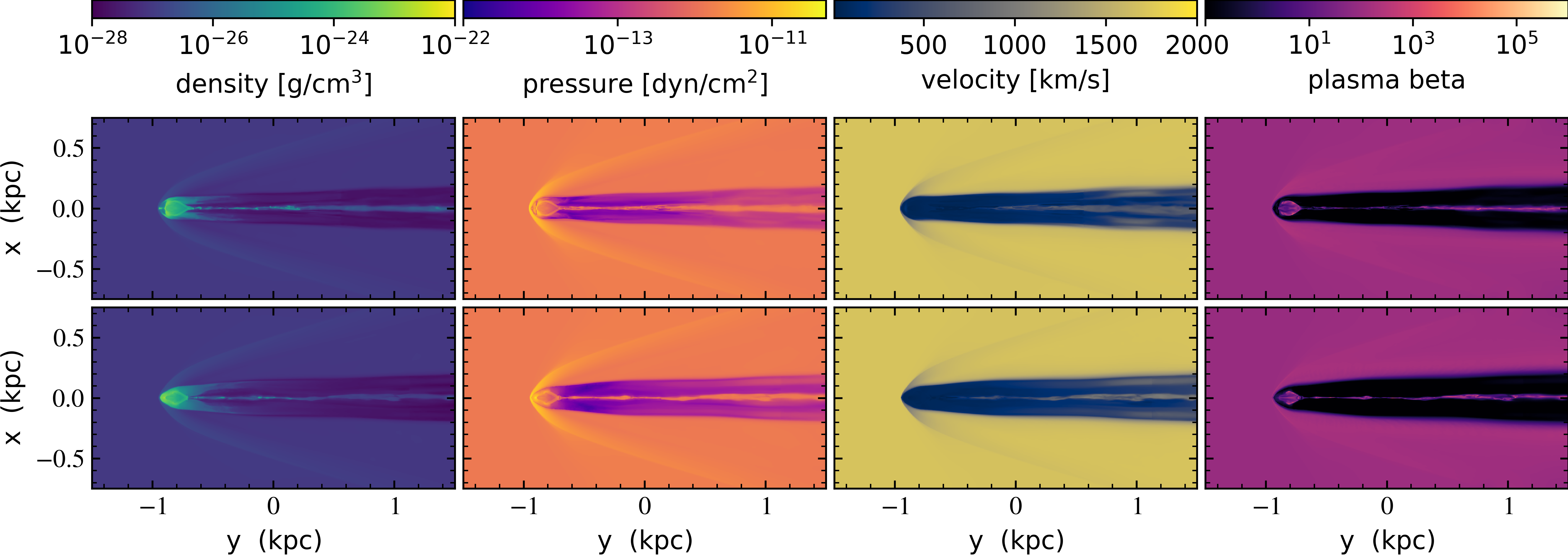}
\caption{Slices of important quantities for runs M3.5-trans-aniso with 2 (top) and 1 (bottom) levels or refinement at $4 t_{\rm cc}$.}
\label{fig:overview}
\end{figure}

In order to test for numerical convergence we have repeated our M3.5-trans-aniso run adopting only one level of additional refinement, as opposed to two in the cases described above. Runs with higher spatial resolution than the production runs presented here are prohibitively expensive. Results for the mass evolution are shown in Fig.~\ref{fig:convergence}. Slices of important quantities are shown in Fig.~\ref{fig:overview}. We find that our results are remarkably similar between the runs with 1 and 2 levels of refinement. Interestingly, the convergence is better than in runs without thermal conduction, as the diffusivity injected by the thermal conduction supersedes numerical effects.

\end{document}